\documentclass[aps,pre,preprint]{revtex4-2}

\usepackage{graphicx}
\usepackage{dcolumn}
\usepackage{bm}
\usepackage{xcolor}
\usepackage{amsmath}
\usepackage{amsfonts}
\usepackage{tikz}
\usepackage{pgfplots}
\pgfplotsset{compat=1.18}
\usepackage{multirow}

\begin{document}
\preprint{APS/123-QED}
\title{\textbf{Complex Pressure Node Formation and Resonances Induced by Scatterers in a Standing-Wave Acoustic Cavity}} 
\author{Rizwan Ullah\textsuperscript{a}, Andres Barrio-Zhang\textsuperscript{a}, Arezoo M. Ardekani\textsuperscript{a,*}}
\affiliation{\textsuperscript{a}School of Mechanical Engineering, Purdue University, West Lafayette, 47907, IN, United States}
\email{\textsuperscript{}Corresponding author: ardekani@purdue.edu}

\begin{abstract}
Acoustic pressure nodes in acoustophoretic devices are crucial for applications in tissue engineering, cell analysis, and particle trapping. Typically, a single primary node forms at the half-wavelength resonance condition, while its shape and position are constrained by the dimensions of the channel. The generation of additional nodes, along with control over their positions and shapes, is highly desirable in biomedical applications and could significantly enhance particle manipulation capabilities. To explore this potential, we numerically demonstrate the formation of additional, complex-shaped, nodes alongside the primary one, by a circular scatterer within a rectangular cavity. We identify three distinct types: ring, protuberant, and crescent nodes, whose formation depends on the size of the scatterer, its placement in the middle of the channel, and its corresponding resonant frequency. The key mechanism behind this formation is the enhancement of internal reflections, increasing destructive interference to promote node generation. To achieve this, we employ three key concepts: using a low-aspect ratio channel, positioning a rigid circular scatterer in the middle of the channel, and modeling all surfaces as perfect reflectors. Furthermore, we analyze the impact of the scatterer on acoustic pressure and quality factor, which is defined as the ratio of stored acoustic energy to damped acoustic energy per cycle.
We show that additional nodes emerge in the presence of a scatterer, but they come at the cost of reduced acoustic pressure and energy in the channel. In summary, this study provides information on generating complex nodes in a standing wave acoustic cavity at the fundamental frequency, which has numerous applications for particle manipulation in acoustofluidic devices.
\end{abstract}

\maketitle
\section{Introduction}
Acoustofluidic devices have been extensively used in various domains, including biomedical, chemical, medical, and environmental fields, for diverse applications. These include mixing~\cite{Mixing}, drug delivery~\cite{Drugdelivery}, biosensing~\cite{dropletmanipulation}, sample preparation~\cite{SamplePreparation}, non-invasive particle or cell analysis~\cite{cellanalysis}, separation~\cite{Separation}, patterning ~\cite{changingparadigm_tissueengineering,cellpatternin_muscelgrowth}, sorting~\cite{sorting}, particle trapping~\cite{trapping_LateralTransducerModes,nervegrowth,cellpatternin_muscelgrowth,muscle_growth,cellTrapApplication_tissueEngineering,hammarstrom_seed_2012,barrio-zhang_acoustically_2024}, concentrating~\cite{Concentration,nervegrowth,cellpatternin_muscelgrowth,muscle_growth,cellTrapApplication_tissueEngineering}, droplet manipulation~\cite{dropletmanipulation}, and tissue engineering~\cite{changingparadigm_tissueengineering, nervegrowth,cellpatternin_muscelgrowth,muscle_growth,cellTrapApplication_tissueEngineering}. These applications benefit greatly from the formation of pressure nodes, as these nodes act as stable locations where the forces on the particles or cells are balanced~\cite{BrussModel,Bruss2014numerical}. Control over the shape and position of these nodes is important for these applications~\cite{controlOverNodalPosition}. For example, in tissue engineering, these nodes help arrange cells in specific patterns that mimic the natural structure of tissues~\cite{tissueEngineeringbook}. Straight or grid-like patterns are beneficial for guiding nerve growth~\cite{nervegrowth} and forming blood vessels~\cite{angiogenesis}, essential for creating functional networks. Pressure nodes also make it possible to build complex tissue structures, such as layered skin, blood vessels, or aligned muscle fibers, by precisely organizing cells in the right shapes and arrangements~\cite{nervegrowth,cellpatternin_muscelgrowth,muscle_growth,cellTrapApplication_tissueEngineering}. For example, linear cell patterns have been used to increase the tensile strength of tissues~\cite{muscle_growth} and to facilitate the formation of blood vessels~\cite{angiogenesis}, while hexagonal patterns have been used to enhance adhesion at cell-cell interfaces~\cite{cell-cell_adhesion}. 

Acoustic devices designed to create these nodes are mainly categorized into bulk acoustic wave (BAW) devices and surface acoustic wave (SAW) devices. SAW devices are commonly used to generate linear nodes, whereas BAW devices are more frequently used to form complex node patterns due to their larger acoustic domains compared to SAW devices~\cite{changingparadigm_tissueengineering}. In both types, the nodal position is typically determined by the choice of the channel geometry in the direction of wave propagation~\cite{Thinreflector}, as well as the shape of the transducers and interdigital transducers (IDTs)~\cite{changingparadigm_tissueengineering}. Once the device is fabricated, there is limited flexibility in adjusting the position of the nodes, such as operating at different harmonic intervals, but there is typically no flexibility in altering its shape. 

Various strategies have been used to adjust the position of nodes in acoustophoretic devices, including the use of hybrid materials for channel fabrication~\cite{HybridMaterialChannel}, thin reflectors~\cite{Thinreflector}, use of echo channels~\cite{echo-channel}, altering the speed of sound~\cite{AlteringSonicVelocity}, changing driving frequencies~\cite{ChnagingDrivingFrrequency}, multi-wavelength devices~\cite{Multiwavelength1, Multiwavelength2}, and using particle response time alongside higher resonance modes~\cite{ResponseTime_HigherModes}. Although these methods provide some flexibility in nodal positioning, the options are generally limited to a few predefined positions. Moreover, most are based on frequency tuning, which reduces the acoustic energy density and radiation force due to resonance shifts away from the natural frequency~\cite{controlOverNodalPosition}. Additionally, these methods typically generate a single pressure node per driving frequency~\cite{LowerThroughput, HybridMaterialChannel,echo-channel,ChnagingDrivingFrrequency}. However, for particle-focusing applications, single-node devices often become ineffective when the node is fully occupied ~\cite{HigherHarmonics}, especially in suspensions with high particle concentrations~\cite{HigherHarmonics}. Therefore, techniques must be used to increase the nodal capacity to accommodate more particles while also providing flexibility in the nodal locations.

The need to increase nodal capacity to accommodate more particles has been addressed through parallel configurations~\cite{parallelSetup}, the use of higher harmonics~\cite{HigherHarmonics}, or the creation of two-dimensional trapping regions using multiple transducers~\cite{trapping_LateralTransducerModes}. However, higher-harmonic setups often require larger devices to align the resonator dimensions with multiple wavelengths, leading to increased losses and reduced radiation forces due to the enlarged fluid domain. Furthermore, the effects of higher harmonics do not scale linearly~\cite{HigherHarmonics}, limiting their efficiency. Although two-dimensional nodes generated by multiple transducers can improve nodal capacity, they come at the cost of increased heat generation within the acoustofluidic device, which is particularly detrimental when handling biological particles~\cite{baasch2024wholeChannel}. The formation of multiple nodes at the fundamental frequency offers a promising solution to these challenges, addressing the low nodal capacity of single-node setups while eliminating the need for larger devices or additional transducers. Moreover, if the location of these nodes can be adjusted, then they would provide significant advantages for diverse applications, particularly in tissue engineering.

In this paper, we address the aforementioned challenges, specifically: generating additional nodes at the fundamental frequency with control over their locations and showcasing the possibility of generating nodes of complex shapes. For example, we demonstrate the formation of ring nodes using a circular scatterer. This is achieved through a simple concept of maximizing reflection in the acoustophoretic channel, thereby increasing the likelihood of destructive interference, as pressure nodes result from such interference. To achieve this, we used three key strategies: (1) incorporating a scatterer inside the channel, (2) employing a low-aspect-ratio channel, and (3) modeling the surfaces as perfect reflectors of the waves. The setup is shown in Section II as Fig.~\ref{Low_Aspect_Ratio_Channel}. This approach establishes wave interactions among the top and bottom walls, adjacent perpendicular walls, side walls, and scatterer and walls. These interactions are crucial because they can generate additional nodes, as observed and reported in this study. We numerically modeled the formation of nodes near a circular scatterer within a rectangular channel. The channel used had a length-to-width ratio of four, hereafter referred to as the aspect ratio (AR). This AR is intentionally selected to increase the probability of waves' destructive interactions and is much lower than those typically reported in the literature~\cite{baasch2024wholeChannel,HigherAR_1,HigherAR_2,bruss_twopeaks}. In the conventional higher-AR channels, the side walls are farther apart, leading to negligible wave interactions between adjacent perpendicular walls and side walls, assuming that these channels are two-dimensional with negligible thickness. To further enhance the wave reflection, we positioned a scatterer precisely at the center of the channel, ensuring equal interaction with all the channel walls. All surfaces are modeled with sound-hard wall boundary condition to ensure perfect reflections, thus maximizing the likelihood of destructive interference and facilitating the generation of additional pressure nodes.

The structure of this paper begins with an introduction to the geometric setup of the problem, followed by the governing equations essential to understanding and solving it accurately.  This is followed by the results and discussion section, which highlights not only the creation of additional nodes due to the scatterer but also examines its impact on key performance parameters of the acoustic field--acoustic pressure and quality factor.

\section{Geometric setup}
\begin{figure}[htbp]
    \centering
    \begin{tikzpicture}
    \begin{scope}[shift={(-5,0)}] 
    \shade[top color=red!80, bottom color=white] (-2,0) rectangle (2,0.5); 
    \shade[top color=white, bottom color=blue!80] (-2,-0.5) rectangle (2,0); 
    \draw[thick, black] (-2,-0.5) rectangle (2,0.5); 
    \fill[gray!30] (-2.3,0.5) rectangle (2.3,0.9) node[midway,black] {Silicon chip};
    \fill[black] (-3,0.9) rectangle (3,1.3) node[midway,green!70!black] {Transducer};
    \fill[gray!30] (-2.3,-0.9) rectangle (2.3,-0.5);
    \fill[black] (-3,-1.3) rectangle (3,-0.9) node[midway,green!70!black] {Transducer};
    \fill[gray!30] (-2,-0.5) rectangle (-2.3,0.5);
    \fill[gray!30] (2,-0.5) rectangle (2.3,0.5);
    \filldraw[black] (0,0) circle(0.07); 
    \draw[thin] (-2.15,-0.5) -- (-2.15,-0.2);
    \draw[thin] (-2.15,0.2) -- (-2.15,0.5); 
    \node[left] at (-1.95,0) {\textbf{$w$}}; 
    \draw[thin] (-2,-0.6) -- (-0.3,-0.6); 
    \draw[thin] (0.3,-0.6) -- (2,-0.6);   
    \node[below] at (0,-0.4) {\textbf{$l$}};    
    \end{scope}
    \end{tikzpicture}
   \caption{Top view of the low-aspect-ratio BAW device. This device, with an aspect ratio of \( l/w = 4 \), features a silicon chip (gray) containing a water channel, excited by a piezoelectric transducer (black). The channel exhibits a standing wave in the \( y \)-direction, where the primary node—the region where the acoustic pressure gradient is zero—corresponds to half-wavelength resonance. This primary node is highlighted in white, and the scatterer (black dot) is fixed at its center inside the channel. The red color represents positive pressure, while the blue represents negative pressure.}
    \label{Low_Aspect_Ratio_Channel}
\end{figure}
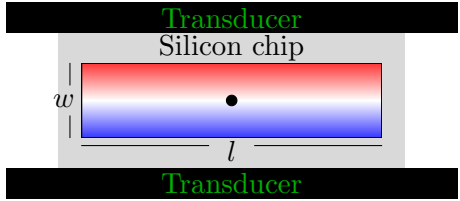

The formation of pressure nodes in an acoustophoretic channel is the result of the destructive interference of sound waves. Therefore, the aim was to maximize the reflection off the surfaces of the channel and that of the scatterer. For this purpose, we selected a channel with a low AR, as shown in Fig.~\ref{Low_Aspect_Ratio_Channel}, so that the reflections interact with each other. The schematic shows the water channel, made on a silicon chip and excited by a piezoelectric transducer, where $w$ represents the width of the water channel and $l$ its length. Fig.~\ref{Low_Aspect_Ratio_Channel} shows the top view of the channel, where the channel thickness is ignored to emulate a shallow channel \cite{Shallow_Channel}.The channel length is \( l = 2\lambda \), while the width is \( w = \lambda/2 \). Here, \( \lambda \) represents the wavelength corresponding to the half-wavelength resonance of the free channel, a channel free from any scatterer. Half-wavelength resonance can be defined as a resonance condition in which the dimension of the channel in the direction of the standing wave is half the wavelength. The pressure node corresponding to half-wavelength resonance in BAW devices typically appears at the location \( w/2 = \lambda/4 \). For example, in our study, \( w = \lambda/2 \), thus satisfying the half-wavelength resonance condition for the generation of a standing wave in the \( y \)-direction. The corresponding fundamental resonance frequency for the standing wave at half-wavelength resonance can be calculated as \( f = \frac{\omega}{2\pi} = \frac{c_0}{\lambda} \). Here, $c_0$ represents the speed of sound and $\omega$ the angular frequency. A rigid circular scatterer is placed in the center of the primary node. The size of the scatterer is varied so that different values of $ka$ are achieved, where $k$ is the wavenumber, defined as \( k = 2\pi/\lambda \), and $a$ is the radius of the scatterer. Here, $ka$ is a dimensionless parameter that indicates whether the scatterer falls within the so-called Rayleigh limit. The value of $ka$ ranges from near zero to $0.4$, covering scatterers ranging in size from nanoscale to microscale. Since the scatterer is circular, the results can also be interpreted for a particle trapped within a node, assuming that it does not move over time.

\section{Governing Equations}
In this section, we present the governing equations that are solved to study the impact of the scatterer on the acoustic field. We utilized the thermoviscous acoustic interface of COMSOL Multiphysicsis version 6.0 \cite{comsol2021} for this purpose, as it efficiently captures thermal and viscous losses in geometries with micro- or nanoscale dimensions~\cite{COMSOLThermoviscousAcoustics}. The dependent variables for this study are pressure ($p$) velocity ($\mathbf{u}$) and temperature ($T$). Since the problem involves sound waves, which are inherently harmonic, it is more convenient to express all the solved equations in the frequency domain. The derivation of these equations begins with the assumption of small harmonic oscillations superimposed on a quiescent background field, defined by ($p_0, \mathbf{u}_0, T_0, \rho_0$). Following Nyborg's perturbation technique \cite{nyborg1965}, the dependent variables can be expressed as:  
\begin{subequations} \label{eq:perturbation}
\begin{align}
p &= p_0 + p_1 + \ldots \label{eq:perturbation_a} \\
\mathbf{u} &= \mathbf{u}_0 + \mathbf{u_1} + \ldots \label{eq:perturbation_b} \\
T &= T_0 + T_1 + \ldots \label{eq:perturbation_c} \\
\rho &= \rho_0 + \rho_1 + \ldots \label{eq:perturbation_d}
\end{align}
\end{subequations}

Here, $p_1$, $\mathbf{u_1}$, $T_1$, and $\rho_1$ represent the first-order perturbations in the pressure, velocity, and temperature fields, respectively, and are assumed to be harmonic in nature. Substituting these equations into the conservation equations for mass, momentum, and energy and neglecting higher-order terms yields the following equations:
\begin{equation}
i \omega \rho = - \rho_0 (\nabla \cdot \mathbf{u})
\end{equation}
\begin{equation}
i \omega \rho_0 \mathbf{u} = \nabla \cdot \left( -p \mathbf{I} + \mu (\nabla \mathbf{u} + (\nabla \mathbf{u})^T) + \left( \mu_B - \frac{2}{3}\mu \right)(\nabla \cdot \mathbf{u}) \mathbf{I} \right)
\end{equation}
\begin{equation}
i \omega \left( \rho_0 C_p T - T_0 \alpha_0 p \right) = - \nabla \cdot (-K \nabla T) + Q
\end{equation}
Finally, the linearized equation of state relates variations in density ($\rho$), pressure ($p$), and temperature ($T$) can be given as:  
\begin{equation}
\rho = \rho_0 (\beta_T p - \alpha_0 T).
\end{equation}

In the aforementioned equations, $\omega$ is the angular frequency, $\mu$ represents the dynamic viscosity, $\mu_B$ denotes the bulk viscosity, $K$ is the thermal conductivity, $C_p$ is the specific heat capacity at constant pressure, $\alpha_0$ is the thermal expansion coefficient, $\beta_T$ is the isothermal compressibility, and $Q$ represents any external heat source, which is zero for this study. For a detailed derivation and discussion on the applicability of the aforementioned equations, the readers are encouraged to refer to \cite{BruusChapter2,COMSOL_Thermoviscous}.

The ultrasound excitation of the boundaries as shown in Fig.~\ref{fig_Actuation} is modeled by the boundary condition defined on the first-order velocity, $v_1$.  
\begin{equation}
\mathbf{n} \cdot v_1 = v_{bc}e^{i\omega t}.
\label{vbc}
\end{equation}
In this formulation, $\mathbf{n}$ denotes the normal unit vector directing outward. The term \( v_{bc} \) represents the magnitude of the harmonic oscillation of the first-order velocity \( v_1 \) on the excited walls of the channel~\cite{BrussModel, bruus2014perturbation}. It can be mathematically expressed as \( v_{bc} = \omega d_0 \), where \( d_0 \) is the magnitude of the displacement of the excited wall due to transducer actuation~\cite{BrussModel, bruus2014perturbation}. All surfaces—the walls of the channel and the surface of the scatterer—are maintained at an isothermal temperature such that
\begin{equation}
T = T_0, \quad \text{on all the surfaces.}
\end{equation}
While the velocity of the surfaces is constrained by the no-slip wall condition:
\begin{equation}
v = 0, \quad \text{on all the surfaces.}
\end{equation}
To model the surfaces as sound-hard walls to maximize the reflection off the surfaces, we imposed the boundary condition as 
\begin{equation}
\mathbf{n} \cdot \nabla p_1 = 0, \quad \text{on all surfaces.}
\label{rigidwall}
\end{equation}
Here, $\nabla p_1$ is the gradient of the first-order pressure field, $p_1$.

When the walls of the channel are excited, the no slip condition on the surfaces of the walls causes viscous and thermal dissipation near the boundaries of the channel and the surface of the scatterer. Thus, the acoustic viscous boundary layer ($\delta$), valid near rigid walls \cite{BruusChapter2}, can be calculated using the following equation, where $\nu$ denotes the kinematic viscosity of the water. 
\begin{equation}
\delta = \sqrt{\frac{2 \nu}{\omega}}
\label{delta}
\end{equation}
We intentionally neglected the effects of thermal losses, as they are approximately seven times smaller than viscous losses in water. The model is solved under standard conditions of $25^\circ \mathrm{C}$ and atmospheric pressure, with the standard properties of water obtained from the COMSOL Material Library.  The results presented here are based on the converged mesh, where all acoustic field parameters exhibit a relative difference of less than \( 10^{-3} \) compared to the reference mesh results. The trend of acoustic field convergence has been found consistent with the mesh convergence studies of \cite{BrussModel,ItalyC(g),Bruss2014numerical, ImpactofChannelHeight,nama2015acoustic}.

\section{\label{sec:level1}Results and discussion\protect}

In this section, we examine the impact of the size of the scatterer on the variation in acoustic pressure, the formation of secondary nodes, and the changes in the quality factor. The size of the scatterer and its corresponding resonance frequency are characterized by the parameter $ka$. Before presenting these results, we first outline the excitation method and the characteristics of the channel. The subsequent subsections then provide a detailed discussion of the findings.

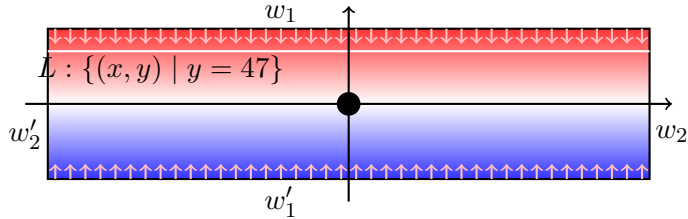
\begin{figure}[htbp] 
\centering
\begin{tikzpicture}
\shade[top color=red!80, bottom color=white] (-4,0) rectangle (4,1); 
\shade[top color=white, bottom color=blue!80] (-4,-1) rectangle (4,0); 
\draw[thick] (-4,-1) rectangle (4,1); 
\filldraw[black] (0,0) circle(0.15); 
\node at (-0.9, 1.2) {$w_1$}; 
\node at (-0.9, -1.3) {$w_1'$}; 
\node at (4.3, -0.4) {$w_2$}; 
\node at (-4.3, -0.4) {$w_2'$}; 
\foreach \x in {-3.9,-3.7,...,3.9} {
    \draw[->, thick, pink] (\x,1) -- (\x,0.8); }
\foreach \x in {-3.9,-3.7,...,3.9} {
    \draw[->, thick, pink] (\x,-1) -- (\x,-0.8);}
\draw[thick, white] (-4,0.7) -- (4,0.7);
\draw[->, thick, black] (-4.3,0) -- (4.3,0); 
\draw[->, thick, black] (0,-1.3) -- (0,1.3); 
\node[black] at (-2.5,0.45) {$L : \{(x, y) \mid y = 47\}$};
\end{tikzpicture}
\caption{Geometric layout of the water channel showing the excitation of the walls $w_1$ and $w_1'$. To model the boundary excitation, we used the method of Muller et al.~\cite{BrussModel}.}
\label{fig_Actuation}
\end{figure}

To model the acoustic field within the fluidic channel in Fig.~\ref{Low_Aspect_Ratio_Channel}, we adopt the strategy of Muller et al.~\cite{BrussModel} while showing the excitation of the fluidic channel independently in Fig.~\ref{fig_Actuation}. The walls \( w_1 \) and \( w_1' \) in Fig.~\ref{fig_Actuation} are excited using the velocity boundary condition given in Eq.~\ref{vbc}, while all other surfaces are modeled as rigid using the hard-wall condition specified in Eq.~\ref{rigidwall}. The actuation of the side walls is performed for two primary reasons. First, the goal is to generate the primary node along the longitudinal direction of the channel. Second, side-wall excitation results in higher efficiency, which means that less thermal dissipation occurs within the channel, which is desirable for applications involving the manipulation of biological particles \cite{baasch2024wholeChannel,SideActuation_higAcousticEnergy}. Moreover, all simulations for the different sizes of scatterers are conducted at their corresponding fundamental resonance frequencies. These frequencies are determined by solving the eigenfrequency for the defined model, focusing on the first eigenmode in the \(y\)-direction. Resonance excitation is crucial for two reasons. First, it ensures maximum energy transfer from the transducer to the fluidic channel. Second, it helps distinguish the effects on the acoustic field caused by factors such as boundary layer dissipation, which result from changes in the size of the scatterer, from those caused by resonance shifts.

\subsection{Impact of the scatterer size on the pressure field}

\begin{figure}[htbp]
    \centering
    \includegraphics[width=1\textwidth]{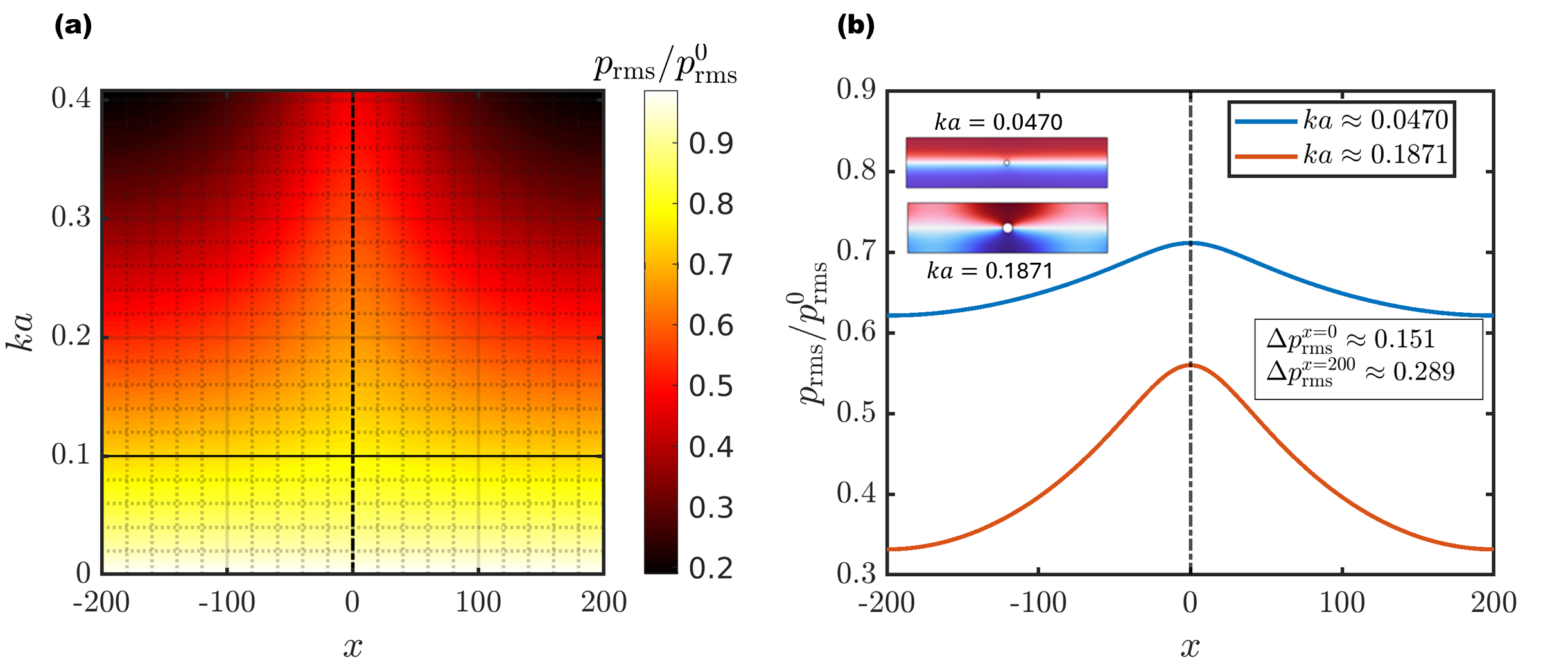} 
    \caption{(a) RMS pressure distribution along the white line $L$ in Fig.~\ref{fig_Actuation}, corresponding to different scatterer sizes and their resonance frequencies characterized by $ka$ (for a channel with $AR=4$). Scatterers in the Rayleigh range ($ka < 0.1$), below the solid black line, exhibit minimal pressure drop, whereas those beyond the Rayleigh range ($ka > 0.1$) experience a significant reduction. The dome-like pattern observed for scatterers with $ka > 0.1$ results from acoustic directivity, where acoustic energy concentrates near the scatterer while decreases near the edges. (b) This plot is specifically intended to highlight the dome pattern by evaluating the RMS pressure along line \( L \) in Fig.~\ref{fig_Actuation} for two scatterers (\( ka = 0.0470 \), \( a = 1.5 \,\mu m \), \( f = 7.4718 \) MHz, and \( ka = 0.1871 \), \( a = 6 \,\mu m \), \( f = 7.4276 \) MHz). 
As $ka$ increases, the pressure distribution becomes more pronounced, with acoustic energy concentrated near the center and causing pressure drop near the edges.}
    \label{fig_pressure(rms)}
\end{figure}

The impact of the scatterer on the acoustic field can be effectively analyzed through the first-order acoustic pressure field, $p_1$. This pressure field directly influences key parameters, such as the acoustic energy ($E_{ac}$) and the radiation force. Both parameters depend on the root mean square pressure (rms), $p_{rms}$ \cite{BrussModel, AcousticBook}. Furthermore, $p_{rms}$ represents the physical field, excluding any imaginary components. For this reason, $p_{rms}$ is reported in Fig.~\ref{fig_pressure(rms)} rather than the instantaneous pressure field. Mathematically, $p_{rms}$ can be given as $p_{rms} = \sqrt{\frac{p_1 p_1^*}{2}}$, where the asterisk denotes the complex conjugate. 

Fig.~\ref{fig_pressure(rms)}(a) shows the pattern of $p_{rms}$ in the presence of various size scatterers represented by $ka$. In a free channel, which is a channel without any scatterer, the pressure distribution of a standing wave at half-wavelength resonance is asymmetric along the wave's direction relative to the node. However, it remains symmetric across the centerline in the direction perpendicular to the standing wave \cite{BrussModel}. In our case, for example, the pressure is asymmetric across the line $(x, 0)$ and symmetric across $(0, y)$, as can be seen in Fig.~\ref{fig_Actuation}. Using this asymmetry, we chose to evaluate the pressure along the line $L : \{(x, y) \mid y = 47\}$ of Fig.~\ref{fig_Actuation}. The $p_{rms}$ in the presence of scatterers is normalized by the rms pressure of the free channel, \( p_{rms}^0 \), evaluated on the same line $L$. This line is positioned outside the viscous boundary layer of the side wall $w_1$, while it remains in the far-field region of the scatterer, where $kr \gg 1$ for most of the scatterers. Here, $r$ represents an arbitrary distance from the surface of the scatterer to the point of interest for the pressure measurement. The pressure evaluated along the line $L$ provides insight into the pressure distribution throughout the channel, except in the near-field region of the scatterer where $kr \ll 1$ and additional ring nodes are observed as discussed in Subsection $B$.

In general, as the size of the scatterer increases, the pressure in the channel decreases. This decrease in pressure can not be attributed to the absorption of sound waves at the surface of the scatterer because the surface of the scatterer is modeled as perfect reflector. Furthermore, the pressure reduction cannot be attributed to a shift from resonance since the channel is excited at its corresponding resonance frequencies when the size of the scatterer changes. The possible pressure reduction could be attributed to three reasons: (1) boundary layer losses in the boundary layers of the scatterer, which increase as the scatterer size increases; (2) wave interaction between the scatterer and the rigid walls of the channel, and (3) the directivity of radiation, which defines how focused the radiation is in a specific direction. However, these mechanisms contribute differently to pressure reduction, depending on whether the size of the scatterer falls within the Rayleigh limit (\( ka < 0.1 \)) or beyond the Rayleigh limit (\( ka > 0.1 \)). For scatterers that satisfy the Rayleigh limit, the pressure decrease can be mainly attributed to the boundary layer losses rather than scatterer-wall interactions or directivity. In this regime, the intensity of the scattering of a particle is proportional to the sixth power of the particle size \cite{rayleigh1896theory}. Due to the smaller size of the scatterers, such scatterers radiate minimally and have a negligible impact on the far-field pressure and a lower probability of interacting with wall reflections. Furthermore, for scatterers with $ka < 0.1$, the pressure drop pattern appears relatively flat, lacking the pronounced dome pattern observed for scatterers beyond the Rayleigh limit—$ka > 0.1$—in Fig.~\ref{fig_pressure(rms)}. Therefore, the directivity of the radiation does not contribute to the pressure drop. For scatterers beyond the Rayleigh limit, the pressure drop is influenced by all three factors discussed above. The pressure reduction results from destructive interference between the waves reflected from the walls and the waves scattered by the scatterer, caused by scatterer-wall interactions. With respect to directivity, as the radiation becomes more concentrated in a specific direction, a dome-shaped pattern forms, leading to a pressure drop toward the edges of the channel. This effect is clearly illustrated in Fig.~\ref{fig_pressure(rms)}(b), where the pressure drop $(\Delta p_{rms})$ is evaluated at two distinct locations, $(x = 0$ and $x = 200)$, for two different scatterers. The directivity factor, given by $\frac{2J(ka \sin\theta)}{ka \sin\theta}$, quantifies the degree to which the sound radiation is focused in a particular direction, with higher values indicating a more concentrated intensity~\cite{kinsler2000fundamentals}. Here, $J$ represents the Bessel function of the first kind and $ka \sin\theta$ is its argument. As the scatterer size increases, so does the directivity, making the dome shape more pronounced. This behavior of the scatterer's directivity closely resembles that of a loudspeaker mounted on a rigid wall. The directivity of sound radiation from a loudspeaker increases in the plane perpendicular to the rigid wall as the value of \( ka \) increases~\cite{kinsler2000fundamentals}. Similarly, the radiations reflected off the scatterer is directed more in the direction perpendicular to the primary node as \( ka \) increases. The primary node enforces a rigid boundary condition on the scatterer, just as a wall does for a loudspeaker. The scatterer is analogous to the loudspeaker, while the primary node is analogous to the rigid wall due to the absence of a pressure gradient at the node.

\begin{figure}[htbp]
    \centering
     \includegraphics[width=0.5\textwidth]{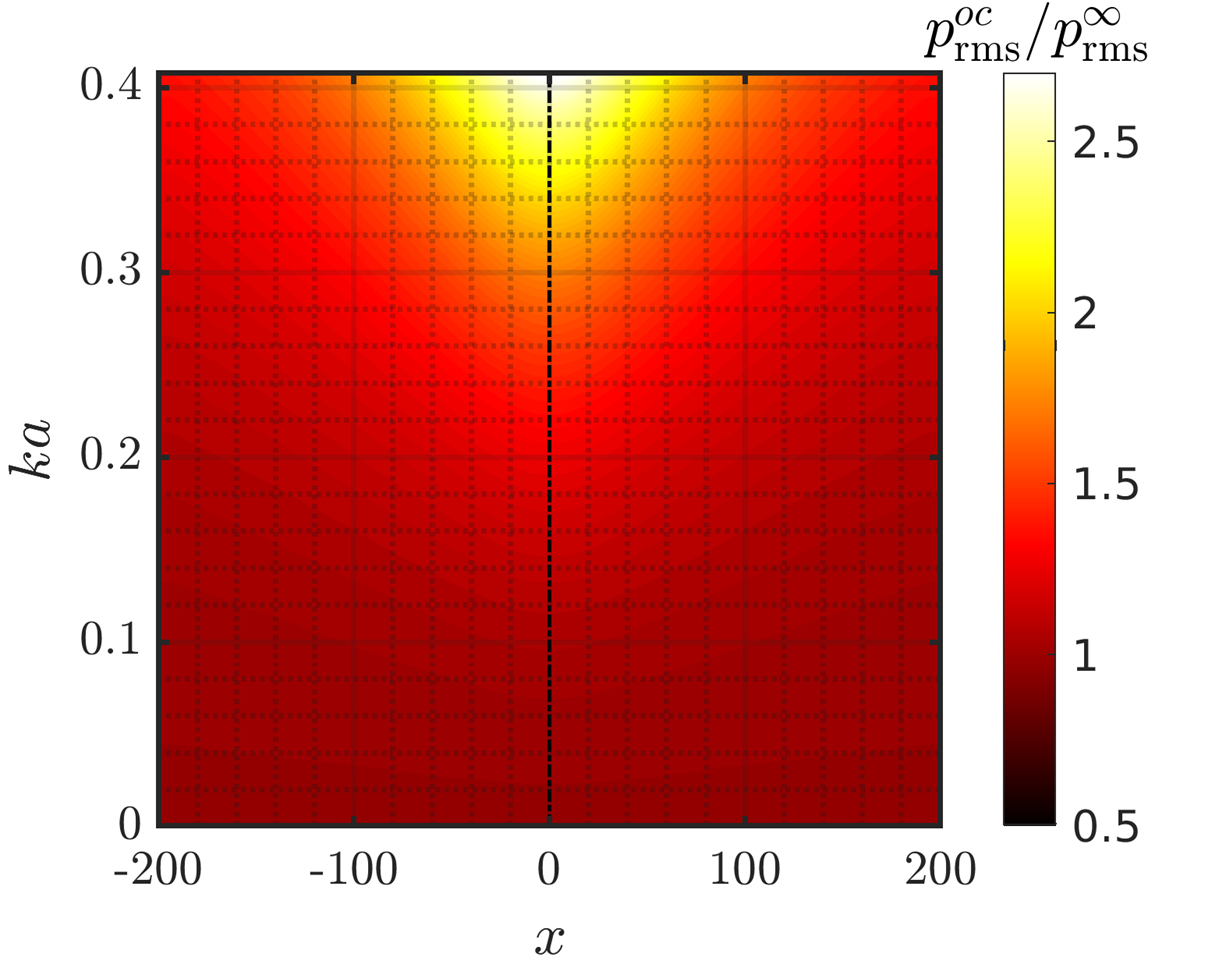}
     \caption{RMS pressure distribution in the presence of scatterers of various sizes, characterized by \( ka \), in an open channel—a channel where the side walls (\( w_2 \) and \( w_2' \)) are far enough that their reflections do not interact with the scattered radiations. The results are evaluated along the line $L$ in Fig.~\ref{fig_Actuation}. The side walls (\( w_2 \) and \( w_2' \)) are modeled using the port boundary condition~\cite{portBC} to mimic an open channel. The results confirm that in the absence of reflections from \( w_2 \), the acoustic pressure increases compared to a free channel due to constructive interference between the scatterer and the reflections from \( w_1 \).}
    \label{fig.scatterer_wall_OC}
\end{figure}

The scatterer in a low-AR channel causes a pressure drop in the channel compared to the free channel, while concentrating the acoustic energy around itself, represented by the dome-shaped pattern of the pressure distribution in Fig.~\ref{fig_pressure(rms)}(a)). To determine whether the scatterer behaves the same way in channels with a larger AR~\cite{baasch2024wholeChannel,HigherAR_1,HigherAR_2,bruss_twopeaks}, we model the side walls \( w_2 \) and \( w_2' \) of the channel as open boundaries using the port boundary condition \cite{portBC}. This condition efficiently mimics channels of larger aspect ratios, which could be defined as a channel where the side walls \( w_2 \) and \( w_2' \) are far enough that their reflections do not interfere with the scattered radiation. The port boundary condition allows all sound energy to exit the channel through these walls, preventing reflections. We refer to such a long channel, modeled with a port boundary condition, as an open channel and present its pressure distribution in Fig.~\ref{fig.scatterer_wall_OC}. The RMS pressure of the open channel, denoted by \( p_{rms}^{oc} \), is normalized by the open-channel pressure in the absence of a scatterer, \( p_{rms}^{\infty} \). Both \( p_{rms}^{oc} \) and \( p_{rms}^{\infty} \) are evaluated along the same reference line \( L \) in Fig.~\ref{fig_Actuation}, on which the results of Fig.~\ref{fig_pressure(rms)} are plotted. Unlike the results in Fig.~\ref{fig_pressure(rms)}, the pressure throughout the channel is greater than unity, indicating that the scatterer improves the pressure in an open channel (see Fig.~\ref{fig.scatterer_wall_OC}). The increase in pressure in the presence of a scatterer in an open channel is mainly due to the absence of out-of-phase reflections from the side walls \( w_2 \) and \( w_2' \), while the wave interaction between the scatterer and \( w_1 \) is in phase. The pressure amplification due to the in-phase interaction between \( w_1 \) and the scatterer outweighs the pressure drop caused by the additional viscous losses due to the scatterer. Furthermore, the relatively higher pressure values near the center of the channel in Fig.~\ref{fig.scatterer_wall_OC} compared to the edges of the channel indicate the localization of the acoustic energy around the scatterer. This is due to the same phenomenon of directivity, which was also observed in Fig.~\ref{fig_pressure(rms)} for the low-AR channel. The concentration of acoustic energy around the scatterer, both in an open channel and in low-AR channels, can be leveraged as a useful technique for acoustic energy localization in applications like high-intensity focused ultrasound and acoustic imaging of biomolecules. This paragraph concludes that a scatterer in all AR channels, whether lower (see Fig.\ref{fig_pressure(rms)}(a)) or higher (see Fig.\ref{fig.scatterer_wall_OC}), concentrates acoustic energy around itself, reflected from the dome-shaped pattern of the pressure distribution. However, the scatterer increases the pressure in channels with higher aspect ratios while reducing it in channels with lower aspect ratios compared to the free channel pressures.

\subsection{Secondary nodes formation}
\begin{figure}[htbp]
    \centering
    \includegraphics[width=0.5\textwidth]{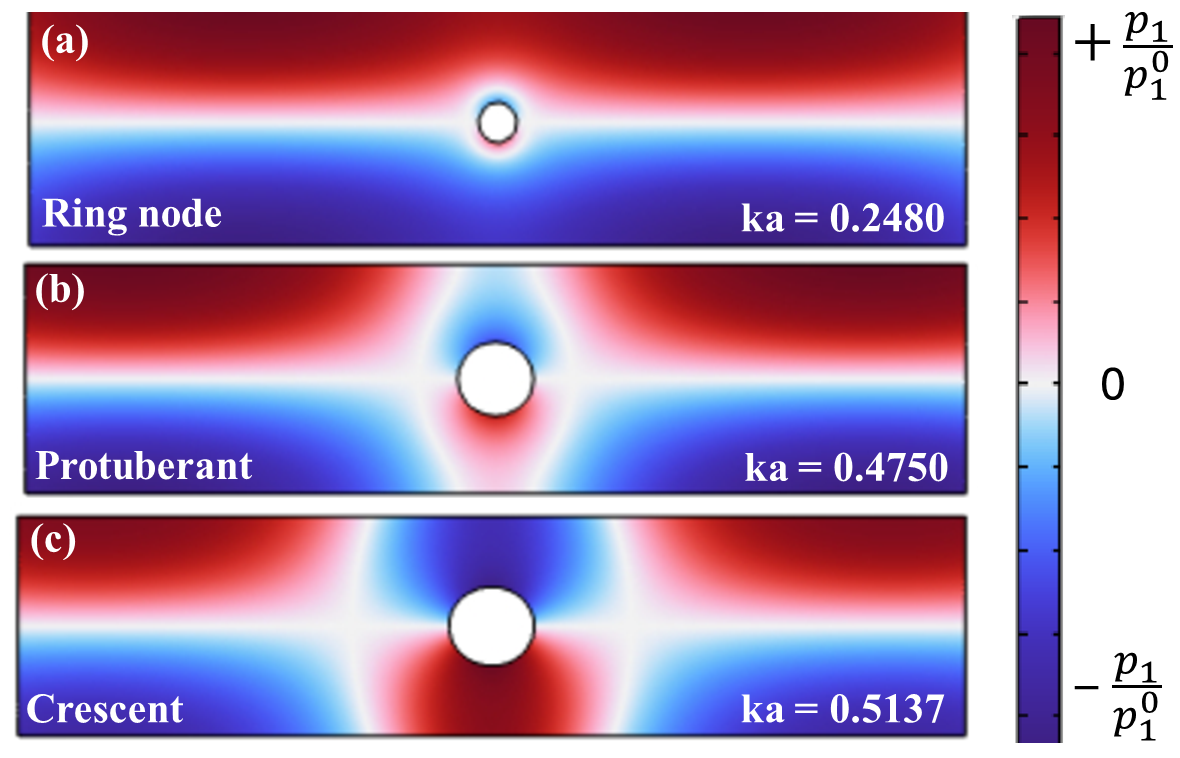}
    \caption{Formation of additional pressure nodes at the fundamental frequency for different scatterer sizes in a channel with \( AR=4 \). (a) Ring node at \( ka=0.2480 \) (\( f=7.3873 \) MHz, \( a=8 \) µm), where the normalized pressure, relative to the free-channel maximum pressure \( p_1^{0} \), reaches a maximum value of \( p_1/p_1^{0} = 0.210 \). (b) Protuberant node at \( ka=0.4750 \) (\( f=7.0729 \) MHz, \( a=16 \) µm), with a maximum pressure of \( p_1/p_1^{0} = 0.116 \). (c) Crescent node at \( ka=0.5137 \) (\( f=6.9933 \) MHz, \( a=17.5 \) µm), where the maximum pressure reaches \( p_1/p_1^{0} = 0.106 \). These distinct node formations arise at different values of \( ka \), highlighting their dependence on scatterer size and its interaction with the background acoustic field.}
    \label{fig.nodes}
\end{figure}

This section aims to report the generation of additional nodes at the fundamental frequency, which is the core idea of this study. We name them secondary nodes to distinguish them from the primary nodes given by the channel dimensions. These secondary nodes exhibit different shapes--ring, protuberant, and crescent--as illustrated in Fig.~\ref{fig.nodes}. The formation of these nodes is determined by the values of $ka$. For $ka < 0.2938$, ring nodes are observed, while for $ka > 0.2938$, the nodes transition into protuberant and crescent shapes.

\begin{figure}[htbp]
    \centering
    \includegraphics[width=0.5\textwidth]{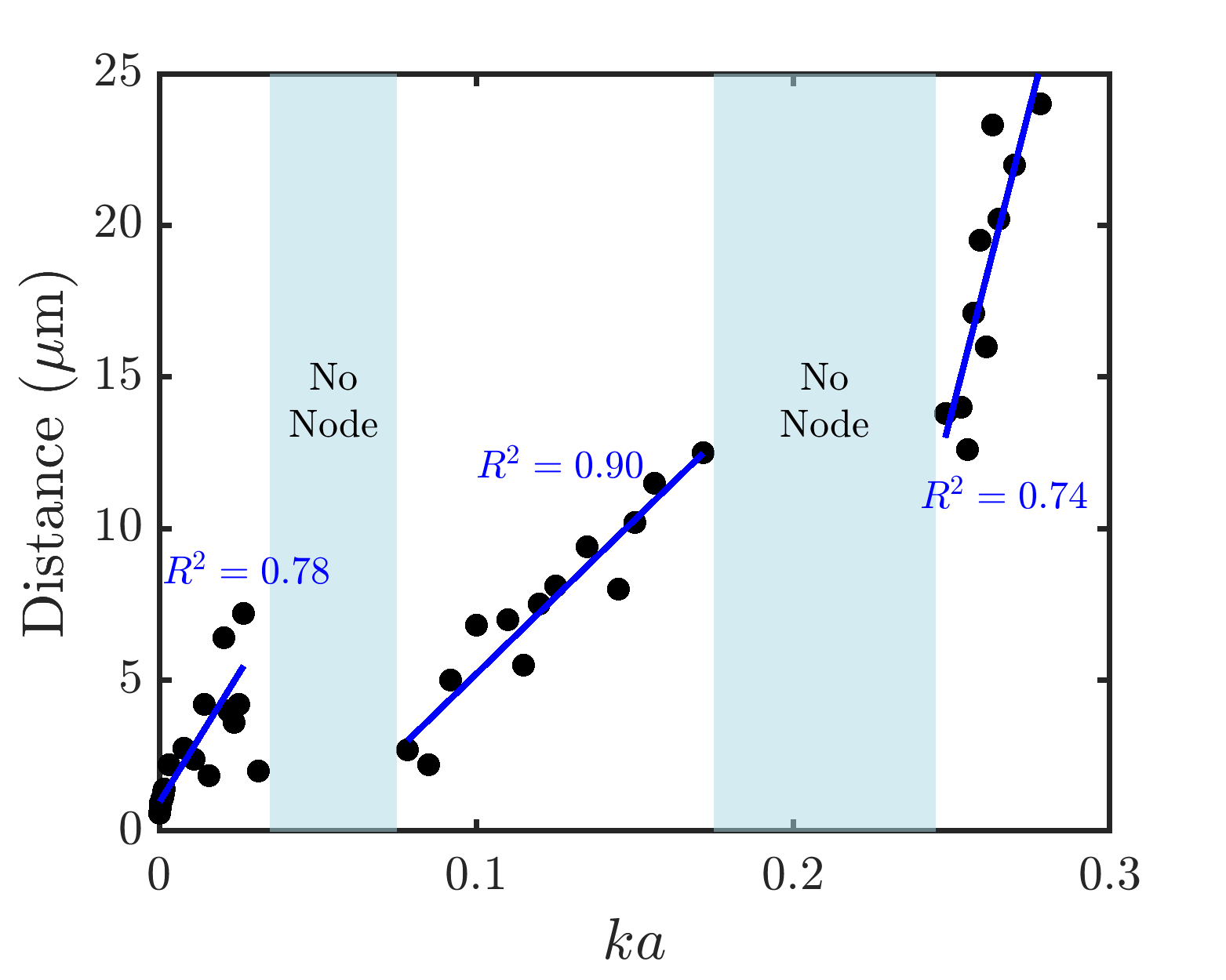}
    \caption{Plot showing the distance of ring nodes from the surface of the scatterer as a function of $ka$. Although the boundaries are modeled as rigid, scatterers of certain sizes—highlighted in light blue—do not reflect out of phase. As a result, node formation does not occur in these regions.}
    \label{fig.RingNodesDistance}
\end{figure}

Ring nodes have been observed to form around both nanoscale and microscale scatterers. Although the region $ka < 0.2938$ produces ring nodes, certain size of the scatterers do not produce these nodes and are shown in Fig.~\ref{fig.RingNodesDistance}. In general, the distance between these nodes and the surface of the scatterer increases with the size of the scatterer (Fig.~\ref{fig.RingNodesDistance}). Ring node formation occurs within the near-field region where $kr<1$. These nodes result from destructive interference between incident planar waves and cylindrical waves reflected off the surface of the scatterers. Their formation depends on two factors--phase shift and mode conversion. Mode conversion refers to the transformation of one wave form into another, for example, in our case, the conversion of planar waves into circular waves upon interaction with the surface of the scatterer. In this study, the $ka$ values that resulted in mode conversion and gradual out-of-phase shift led to the formation of the ring nodes. On the other hand, the values of $ka$ that caused in-phase reflections, preventing node formation. Although mode conversion generally depends on frequency and scatterer size \cite{ModeConversion_ka}, it can also be influenced by the confinement effects of the acoustophoretic channel.

\begin{figure}[htbp]
    \centering
    \includegraphics[width=0.5\textwidth]{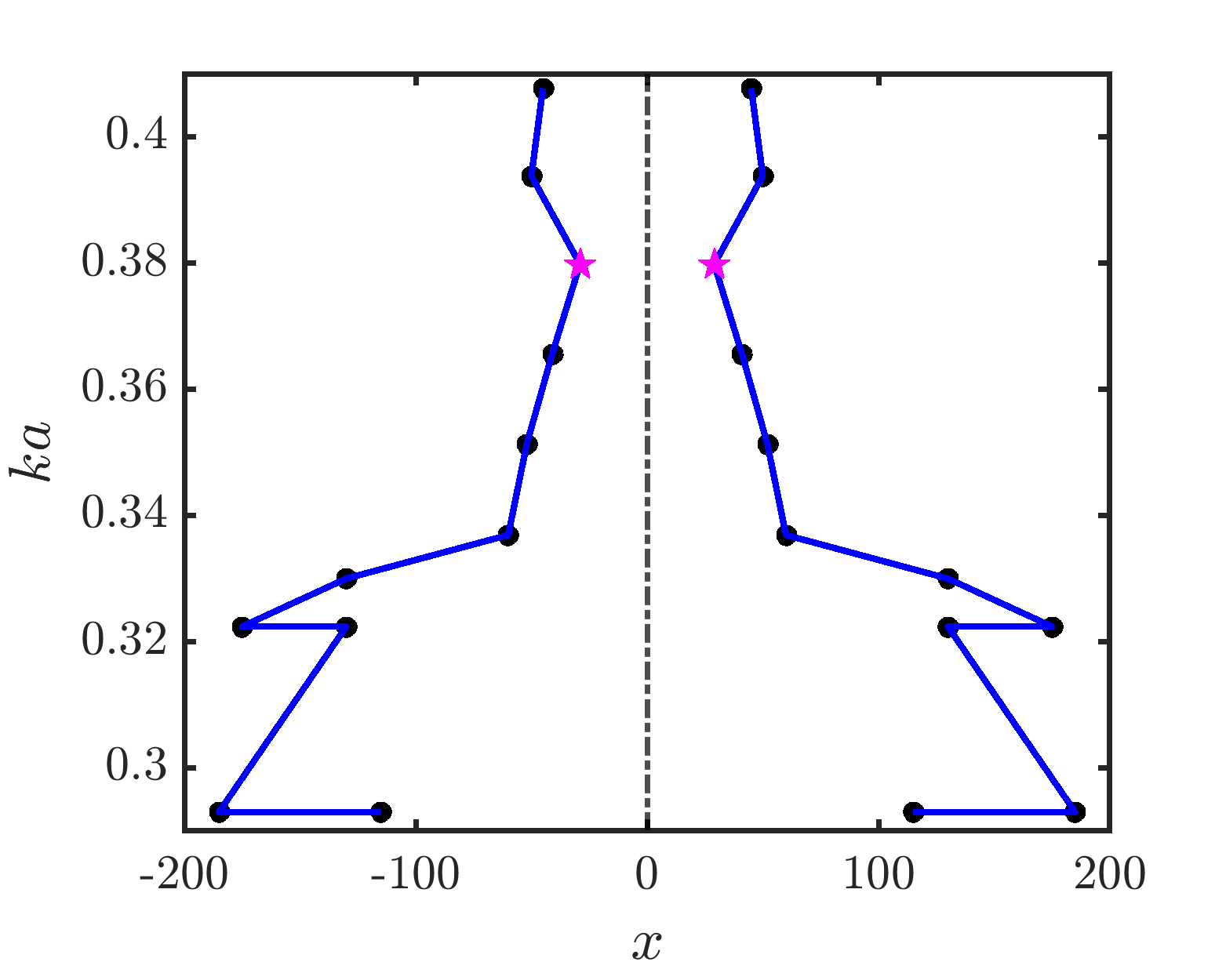}
    \caption{Plot showing the distance of a protuberant node at \( ka=0.3798 \) (\( f=7.2391 \) MHz, \( a=12.5 \) µm), highlighted with a pink star, and crescent nodes from the surface of the scatterer as a function of \( ka \). These nodes appear in pairs, and their distances are measured along the white line \( L \) in Fig.~\ref{fig_Actuation}.}
    \label{fig.CrescentNodesDistance}
\end{figure}

The protuberant and crescent nodes have been observed correspond to the condition where $ka > 0.2930$. These nodes begin to appear when the size of the scatterer satisfies $kr_{w1} \leq 1.28$, where $r_{w1}$ is the distance of the scatterer's surface from the wall $w_1$. Scatterers that cause the transition from ring to crescent nodes initially result in two pairs of crescent nodes, as observed for $ka \approx 0.295$ in Fig.~\ref{fig.CrescentNodesDistance}. This configuration quickly transitions into one pair of crescent nodes as the values of $ka$ increase. These nodes tend to appear near the side walls ($w_2$ and $w_2'$) initially and then begin to travel towards the surface of the scatterer. However, the location of these nodes in the channel changes abruptly without any regular pattern.  For all scatterers that satisfy $kr_{w1} \leq 1.28$, gradual mode conversion is observed along with phase shifts and is the reason for the formation of these nodes (see Figs.~\ref{fig.nodes}(b) and (c)). It should be noted that protuberant nodes do not occur frequently. For values of \( ka \) above 0.293, such nodes are observed only at two specific values of \( ka \). For example, a protuberant node against \( ka = 0.3798 \) is highlighted with a star (pink) in Fig.~\ref{fig.CrescentNodesDistance}.

The formation of additional nodes occurs only in low-AR channels, where wall reflections from walls $w_2$ and $w_2'$ can interact with waves scattered by scatterers. These nodes can appear at the fundamental frequency in low-AR channels but at the expense of pressure reduction in the channel, as shown in Fig.~\ref{fig_pressure(rms)}. In addition, secondary nodes do not form if the channel is sufficiently long, where the scattered waves from the surface of the scatterer cannot interact with the reflected waves from the walls $w_2$ and $w_2'$. However, in such cases, the scatterer can enhance the acoustic pressure due to the amplification of the pressure field resulting from the in-phase reflection of the scattered waves and the waves reflected from walls \( w_1 \) and \( w_1' \) (Fig.~\ref{fig.scatterer_wall_OC}).

\subsection{Impact on Quality Factor}
\begin{figure}[htbp]
    \centering
    \includegraphics[width=1\textwidth]{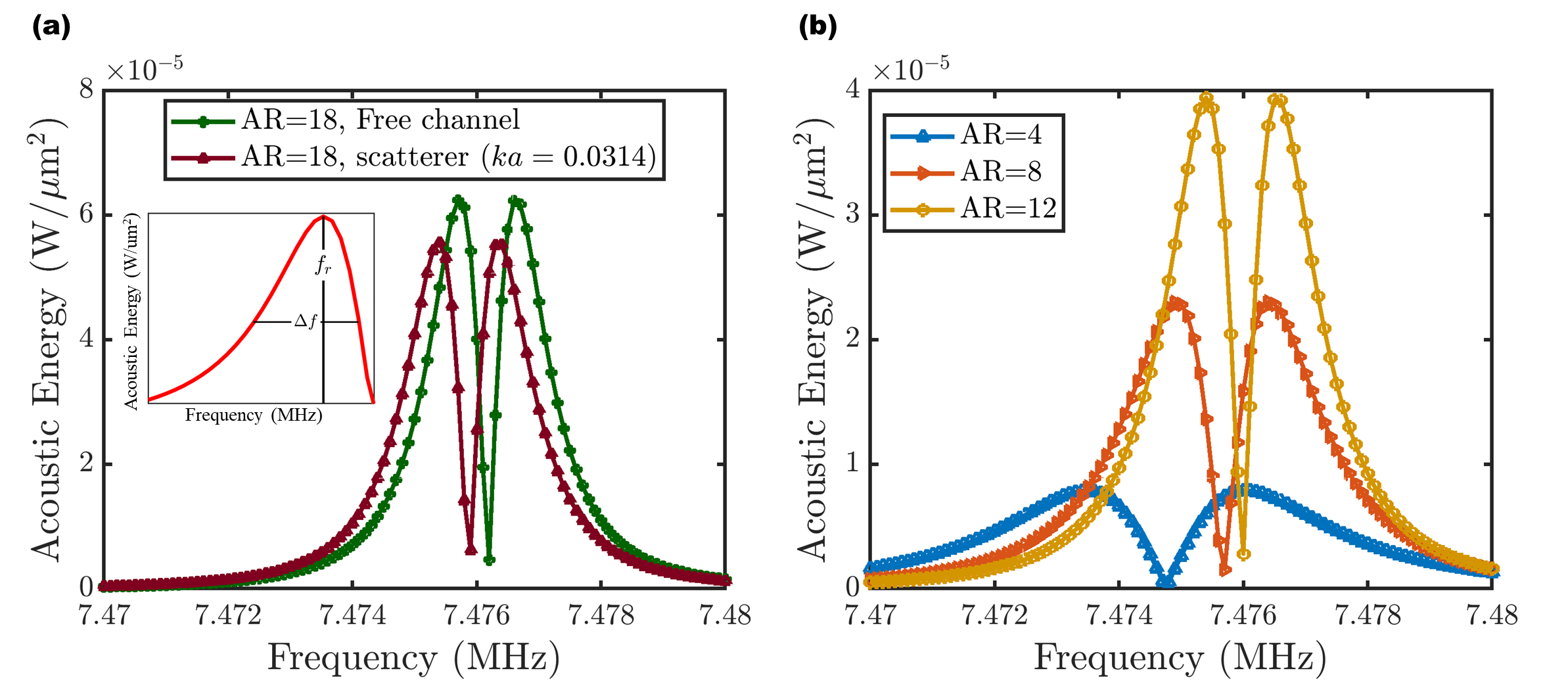}
    \caption{(a) Plot illustrating the effect of a scatterer, defined by \( ka = 0.0314 \) (\( f = 7.4733 \) MHz, \( a = 1 \,\mu m \)), on the quality factor of an acoustophoretic channel with an aspect ratio of 18. The comparison is made between a free channel (without a scatterer) and the same channel in the presence of a scatterer with \( ka = 0.0314 \). The presence of the scatterer reduces the quality factor by lowering the resonance frequency $f_r$ and broadening the frequency range $\Delta f$. (b) Quality factor variation for free channels with three different aspect ratios demonstrates that the quality factor increases as the aspect ratio increases.}
    \label{Fig:QualityFactor}
\end{figure}

The pressure variations caused by changes in the size of the scatterer ($ka$) in Section (IV.A) suggest that the scatterer can also influence the total acoustic energy. Acoustic energy is a critical parameter for evaluating the performance of an acoustophoretic channel, as it directly affects the acoustophoretic efficiency for a given electrical power input~\cite{baasch2024wholeChannel,bruus_scallingLaws}. In addition, it helps determine the channel quality factor. A higher acoustic energy corresponds to a higher quality factor, which is required for efficient particle manipulation in acoustic resonators. Therefore, this section aims to analyze the acoustic energy and the corresponding quality factor in the presence of a scatterer, as illustrated in Fig.~\ref{Fig:QualityFactor}. For comparison, we include the acoustic energy for free channels with different AR values.

The acoustic energy, given by \( E_{ac} = \int I \, dA \), represents the integral of the acoustic intensity (\(I\)) over the entire surface area of the channel, \(A\). The acoustic intensity is calculated using \( I = \frac{1}{2} \Re\{p \mathbf{v}^*\} \), where \(p\) and \(\mathbf{v}\) denote instantaneous acoustic pressure and velocity, respectively, and \(\Re\) refers to the real part of the quantity \(p\mathbf{v}^*\). 
The quality factor (\(Q\)) is calculated as \( Q = \frac{f_r}{\Delta f} \), where \(f_r\) is the resonance frequency, and \(\Delta f\) represents the bandwidth at half the maximum peak amplitude, as shown in Fig.~\ref{Fig:QualityFactor}(a). 

In Fig.~\ref{Fig:QualityFactor}(a), it is evident that for a given channel dimension (tested at AR = 18), the presence of a scatterer reduces the quality factor, making it less suitable for applications requiring strong acoustic resonance at a specific frequency, such as particle trapping. However, this reduction can be advantageous in applications where rapid damping of acoustic energy is desired, such as high-intensity focused ultrasound therapy, or in applications operating across a broader frequency range, such as frequency-based particle separation. Additionally, Fig.~\ref{Fig:QualityFactor}(a) provides insight into how particle trapping influences the quality factor and the performance of an acoustophoretic device. When particles become trapped in nodal positions and remain there over time, the quality factor may be affected. Similarly, in continuous-flow conditions, as one particle moves, another may take its place, potentially altering the device's performance.

\begin{figure}[htbp]
    \centering
    \includegraphics[width=0.5\textwidth]{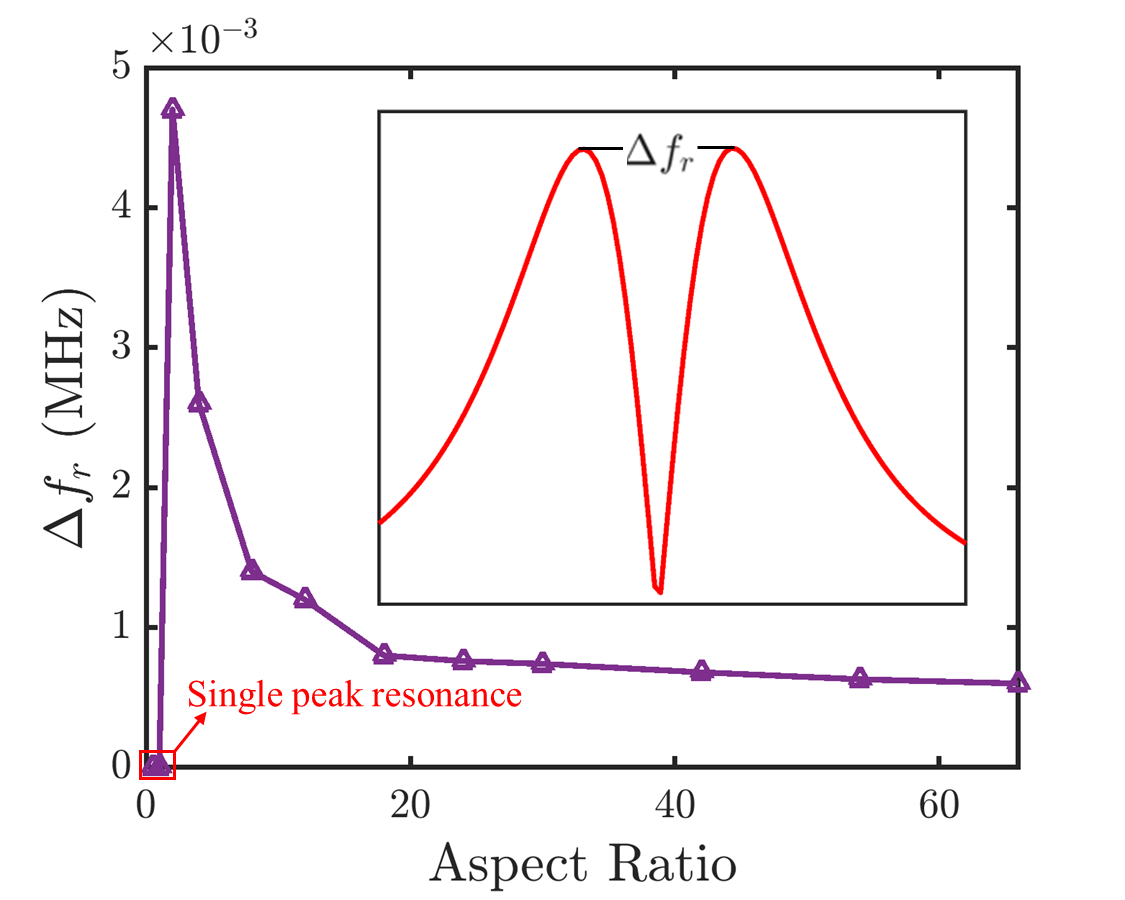}
    \caption{Dependence of the number of resonance peaks on the aspect ratio. A total of 12 free-channel cases are simulated up to an aspect ratio of 66, showing that for aspect ratios $\leq 1$, a single resonance peak is observed, whereas for aspect ratios $> 1$, two resonance peaks emerge.}
    \label{Fig:peaksdistance}
\end{figure}

The data in Fig.~\ref{Fig:QualityFactor}, for both free channels and those containing a scatterer, reveal a two-peak resonance. This two-peak resonance is a function of AR, which is defined as the ratio of length to width. The length and width directions of the channel are shown in Fig.~\ref{Low_Aspect_Ratio_Channel}. If $AR < 1$, a single-peak resonance appears, as shown in Fig.~\ref{Fig:peaksdistance}, resulting in $\Delta f_r = 0$, where $\Delta f_r$ is the frequency difference between the two resonance peaks. However, for all aspect ratios that satisfy $AR > 1$, two resonance peaks emerge, with the peaks moving closer together as the AR increases. At a certain AR value, such as AR = 18 in our case, the decrease in $\Delta f_r$ becomes almost negligible, resulting in a nearly flat trend. These two-peak resonances have been previously reported in experimental studies over a short frequency range, but were fitted with a single Lorentzian curve~\cite{baasch2024wholeChannel,bruss_twopeaks}. Furthermore, the two resonances exhibit symmetry, as shown in Fig.~\ref{Fig:QualityFactor}(a) and (b).

Interestingly, the quality factor increases as AR increases (Fig.~\ref{Fig:QualityFactor}(b)), highlighting the need for channel optimization not only in the height and width directions but also in the length direction, an aspect largely overlooked in many studies~\cite{li2023sensitivity,ImpactofChannelHeight,ItalyC(g), BrussModel}. Studies that considered two-dimensional optimization of the fluidic channel~\cite{li2023sensitivity,ImpactofChannelHeight,ItalyC(g), BrussModel} typically focused on optimizing its cross-sectional area, which can be interpreted as height multiplied by width, while assuming that the third dimension, length, is significantly larger than the other two. As a result, these studies overlooked the influence of the channel's length on the quality factor. Unlike these studies, our two-dimensional study focuses on the width and length directions of the channel, assuming a shallow channel~\cite{Shallow_Channel} in the height direction. Taking into account the length of the channel, which is reflected in the AR of the channel, the results in Fig.~\ref{Fig:QualityFactor}(b) show that the length of the channel significantly affects the quality factor and, consequently, the performance of the acoustophoretic device. The increase in the quality factor as the channel length increases is likely due to the reduction of destructive interference caused by the side walls, $w_2$ and $w_2'$, as they move farther apart. Another possible reason is the stronger coupling of the boundary excitation, modeled through Eq.~\ref{vbc}, to the resonance mode as AR increases. The increase in the quality factor for larger ARs is promising as it indicates improved particle manipulation capabilities of acoustophoretic devices because of the greater storage of acoustic energy. In addition, larger ARs can provide greater nodal capacity for particles, as the length of the primary node increases with the length of the channel. Thus, the combination of a higher quality factor and enhanced nodal capacity in larger AR channels suggests improved channel performance.

\section{conclusion}
This paper reports on the formation of complex pressure nodes at the fundamental frequency within a standing-wave resonance cavity. The approach involves enhancing the wave reflection inside the cavity by modeling all channel surfaces as perfect reflectors while introducing a circular scatterer inside the middle of the channel to maximize the probability of destructive interference, which is responsible for pressure node formation. Our study reveals three distinct node shapes: ring, protuberant, and crescent nodes, which emerge as the size of the scatterer varies and the channel is excited at its corresponding resonance frequency, denoted by $ka$. Here, $k$ is the wavenumber and $a$ is the radius of the scatterer. Ring nodes appear in the near-field region, where interference occurs between cylindrical waves reflected from the scatterer and incident planar waves. In contrast, protuberant and crescent nodes develop when the scatterer satisfies the condition $kr_{w} \leq 1.28$, where $r_{w}$ represents the distance from the surface of the scatterer to the nearest wall. 

As $ka$ increases, the distance of the ring nodes from the surface of the scatterer increases. However, there is a nonmonotonic pattern for the protuberant and crescent nodes. To further analyze the impact of the size of the scatterer on the acoustic field, we examine its influence on the pressure field, the acoustic energy, and the quality factor. Although a scatterer introduces additional complex pressure nodes, it also leads to a drop in acoustic pressure. For scatterers that fall within the Rayleigh limit ($ka < 0.1$), this pressure drop is primarily due to additional boundary layer losses at the scatterer’s surface. However, beyond this limit, the losses arise from both boundary layer effects and out-of-phase interactions between the scatterer and the channel wall, along with increased directivity. Moreover, as $ka$ increases, the acoustic pressure tends to concentrate around the scatterer while decreasing near the channel edges, a phenomenon attributed to the directional focus of the acoustic energy (directivity). This directivity causes an uneven distribution of the acoustic field within the channel, which could be beneficial for particle sorting based on size.

This study also investigates the impact of a scatterer on the quality factor of a channel. The presence of a scatterer reduces the quality factor by broadening the frequency bandwidth, which could be beneficial for applications requiring frequency tuning, such as frequency-based particle separation. Additionally, the quality factor increases as the channel's aspect ratio grows, making it a promising approach to enhance nodal capacity for particle manipulation while minimizing acoustic energy damping. For channels with $AR>1$, two symmetric resonance peaks appear within a narrow frequency range of 0.01 MHz, and their experimental validation is the focus of a future study. Furthermore, future research aims to experimentally confirm the formation of additional nodes, specifically ring nodes, for particle focusing. Future research research will also focus on scatterers of different shapes under perfect reflective wall boundary conditions. 

\section*{Author Contribution Statement}
R.U., A.B.Z., and A.M.A. conceived the research idea and designed the study; R.U. performed the simulations and developed the numerical model; R.U., A.B.Z., A.M.A. analyzed the data. R.U. and A.B.Z wrote the manuscript with input from all authors; A.M.A supervised the research; all authors reviewed and approved the final manuscript.

\section*{Data availability Statement}
The data that support the findings of this study are available from the corresponding author upon reasonable request.

\section*{Acknowledgments}
This project was supported by a Fulbright Program grant, sponsored by the Bureau of Educational and Cultural Affairs of the US Department of State and administered by the Institute of International Education. This work was partially supported by the National Science Foundation award CBET-2141404. The authors have no conflict of interest to disclose.

\bibliography{apssamp}% Produces the bibliography via BibTeX.
\end{document}